\documentstyle[12pt]{article}
\newlength{\blength}
\newcommand{\proof}[1]{\vspace{-.05cm}
\begin{list}{\bf
{Proof:}}{\listparindent=\parindent\parsep=0pt
\labelwidth=0cm
\labelsep=\parindent
\addtolength{\labelsep}{-\blength}
\addtolength{\labelsep}{1.2cm}
\itemindent=-\blength
\addtolength{\itemindent}{\parindent}
\leftmargin=1.2cm}
\item
#1~$\Box$\end{list}
\vspace{.0cm}}

\begin{document}
\thispagestyle{empty}
\setcounter{page}{0}
\newtheorem{theorem}{Theorem}
\newtheorem{coro}{Corollary}
\newtheorem{lemma}{Lemma}
\renewcommand{\theequation}{\thesection.\arabic{equation}}
\renewcommand{\thetheorem}{\thesection.\arabic{theorem}}
\renewcommand{\thelemma}{\thesection.\arabic{lemma}}
\renewcommand{\thecoro}{\thesection.\arabic{coro}}
{\hfill{ULB-TH/00-10, VUB/TENA/00/2, hep-th/0004049}}

\vspace{2cm}

\begin{center}
{\bf CHIRAL FORMS AND THEIR DEFORMATIONS }

\vspace{1.4cm}

XAVIER BEKAERT${}^1$, MARC HENNEAUX${}^{1,2}$ and
ALEXANDER SEVRIN${}^3$

\vspace{.1cm}

${}^1${\em Physique Th\'eorique et Math\'ematique, Universit\'e Libre de
Bruxelles,}\\
{\em Campus Plaine C.P. 231, B-1050 Bruxelles, Belgium} \\
${}^2${\em Centro de Estudios Cient\'{\i}ficos, Casilla 1469,
Valdivia, Chile} \\
${}^3${\em Theoretische Natuurkunde, Vrije Universiteit Brussel} \\
{\em Pleinlaan 2, B-1050 Brussel, Belgium} \\
\end{center}
\centerline{\tt xbekaert@ulb.ac.be, henneaux@ulb.ac.be,
asevrin@tena4.vub.ac.be}

\vspace{1cm}

\centerline{ABSTRACT}

\vspace{- 2 mm}  

\begin{quote}\small
We systematically study deformations of chiral forms with
applications to string theory in mind.  To
first order in the coupling constant, this problem can be translated into
the calculation of
the local BRST cohomological group at ghost number zero. 
We completely solve this cohomology and present detailed proofs of
results announced in a previous letter.
In particular, we show that there is no room for non-abelian, local, deformations of a pure system of chiral p-forms.
\end{quote}
\baselineskip18pt
\noindent

\newpage

\section{Introduction}

A chiral $p$-form  $A$ is defined by the equation
\begin{eqnarray}
F={}^*F.
\end{eqnarray}
where $F\equiv dA$ is the corresponding fieldstrength.
{}From this, it is clear that the dimension, $d$, of space-time is given by
$d=2(p+1)$.  Furthermore,
$p$ should be even in the Minkowski case and odd in
the Euclidean case since only in those cases is the square of the
Hodge $*$-operator equal to the identity. 
Throughout this paper we maintain a Minkowski
signature. Chiral $p$-forms naturally appear in string or M-theory. Chiral
bosons are essential in the worldsheet formulation of the heterotic string
and correspond to $p=0$. Chiral two-forms, which, as we will explain further,
constitute the main motivation for the present study, are central in the
description of the M5-brane. Finally, chiral four-forms appear in type IIB
string theory where they signal the presence of D3-branes.

The strongest motivation for studying deformations of chiral forms arises
from the study of coinciding M5-branes. 
The solitonic objects in M theory
(viewed here as eleven
dimensional supergravity) are M2- and M5-branes. 
These soliton solutions break half of the supersymmetries,
reducing them from 32 to 16.  Their effective worldbrane actions
contain therefore 16 Goldstinos,
which correspond to 8 propagating fermionic degrees of
freedom.
This should
be matched by 8 bosonic degrees of freedom. Obvious candidates for the bosonic
degrees
of freedom of a $p$-brane living in $d$ dimensions are the $d-p-1$
transversal positions of the brane. For the M2 brane ($p=2$
and $d=11$), this saturates the number of bosonic degrees of freedom.
For the M5-brane, however, one needs three additional bosonic degrees of freedom. The little group of the
worldvolume theory is $Spin(6)=SU(2)\times SU(2)$, which means we need 
a (3,1) representation of this. This is precisely a chiral two-form in six
dimensions.

In the low energy limit where bulk gravity decouples, a single M5-brane is
described
by a six dimensional $N=(2,0)$ superconformal field theory 
\cite{sw},
\cite{ls}.
Its field content consists of five scalar fields and a single chiral two-form
\footnote{Throughout this paper we ignore the fermionic degrees of freedom
which does not change any of our conclusions.}.
A Lorentz non-covariant
action was constructed in \cite{PS}, \cite{S} and \cite{APPS}. 
A covariant action was
obtained in \cite{PST} and \cite{BLNPST}. The covariant action contains
appropriate extra auxiliary fields and gauge symmetries.
Partial gauge fixing of the
covariant action yields the non-covariant action.

Once $n$ M5-branes coincide, the situation changes. This can be seen by
compactifying one direction on a circle. For small radius, the resulting
theory is weakly coupled type IIA string theory. When the M5-branes are
transversal to the circle, they appear in the type IIA theory as $n$
coinciding NS5-branes. Not much is explicitely known about this system.
However, when the M5-branes are longitudinal to the circle, they emerge as
$n$ coinciding D4-branes. The effective action for such a
system is a $U(n)$ non-Abelian
Born-Infeld action \cite{tseytlin}. Its leading and next to leading terms
are well understood but discussion about the subleading terms
remains \cite{wati}, \cite{faj}. Ignoring higher derivative terms and
focussing on the leading term, one gets that the dynamics of the D4 system
is governed by 5 scalar fields in the adjoint representation of $U(n)$
coupled to a 5-dimensional $U(n)$ gauge theory. Going back to the supergravity
description, this observation suggests the existence
of a non-abelian extension of chiral 2-forms.

Genuine non-abelian extensions of non-chiral $p$-forms, for
$p\geq 2$ have not yet been constructed. Viewing a 2-form as a connection over
loopspace,
one can show that no straightforward non-abelian extension exists
\cite{teitel} (see also \cite{Nepomechie:1983rb}). 
Dropping geometric prejudices, all local deformations
continously connected
to the free action were constructed in \cite{knaepen}. Though both known and
novel
deformations were discovered, none of them had the  required property that the
$p$-form
gauge algebra becomes truly non-abelian.

Turning back to chiral 2-forms, one finds that M-theoretical
considerations indicate that $n$ coinciding M5-branes constitute a highly
unusual physical system. Indeed, the supergravity description of $n$
M5-branes predicts that both the entropy \cite{kt} and the two-point
function for the stress-energy tensor \cite{gk} scale as $n^3$ in the
large $n$ limit. Anomaly considerations lead to a similar behaviour
\cite{hs}, \cite{bft}. So this suggests that a non-abelian extension of
chiral two-forms falls outside the scope of finite dimensional semi-simple
Lie groups as none of those have a dimension growing as fast as $n^3$
(where $n$ would be the dimension of the Cartan sub-algebra).
It has been argued that ``gerbes" could provide the appropriate
mathematical framework \cite{Kalkkinen:1999uz,Hitchin:1999fh}.

In \cite{bhs1}, we announced the result that no local field theory is able
to describe a system of coinciding M5-branes. This result was 
obtained by showing that local deformations of the action cannot modify the
abelian nature of the algebra of the $2$-form gauge
symmetries.  It holds
under the assumption that the deformed action is continuous
in the coupling constant (i.e., possible non-perturbative
``miracles" are not investigated) and reduces, in the limit of vanishing
coupling constant, to the action describing free chiral $2$-forms.
In particular, no assumption was made on the polynomial order (cubic,
quartic ...) of the interaction terms.

In the present paper we
present detailed proofs of that assertion. The techniques used in this
paper can be applied in a straightforward fashion to prove the results in
\cite{bhs2} as well. There, deformations of chiral four-forms in ten dimensions
were analyzed with as conclusion that the only consistent deformation was
the type IIB coupling of the chiral four-form to the NS-NS and the R-R
two-forms familiar from IIB supergravity 
\cite{Schwarz:1983qr,Howe:1984sr}.

The outline of this paper is as follows.  In the next section, we 
review how the problem of consistent couplings can be reformulated
as a cohomological problem \cite{HenneauxBarnich93C,Stasheff:1997fe}.
We then recall the non-covariant formalism for chiral $2$-forms, and
their BRST formulation (sections 3, 4 and 5).  In particular, 
we point out that the BRST differential $s$ naturally splits as the
sum $s = \delta + \gamma$ of simpler building blocks.
After a brief
section in which we recall the so-called ``algebraic Poincar\'e lemma",
which provides an important tool for our investigations, we turn
to the calculation of the BRST cohomology.  First we compute
the cohomology of $\gamma$ (section 7).  Next,  we compute the
cohomology of $\gamma$ modulo $d$, where $d$ is the spacetime
exterior derivative (sections 8 and 9).  In section 10, we compute
the same cohomologies for the other piece involved in $s$,
namely $\delta$.  In section 11, we put together the calculations of
the previous sections to derive the announced result that the
gauge symmetries for a set of free chiral $2$-forms are rigid
and cannot be deformed continuously in the local field theoretical context.
Our paper ends with a short, concluding section.

\section{Constructing consistent couplings as a deformation problem}
\setcounter{equation}{0}
\setcounter{theorem}{0}
\setcounter{lemma}{0}

The theoretical problem of determining consistent interactions
for a given gauge invariant system has a long history.  It has
been formulated in general terms in \cite{Berends:1985rq}
(see also \cite{Wald:1986bj}).  

The equations for the consistent interactions are rather intricate
because they are non linear and involve simultaneously not
only the deformed action, but also the deformed structure
functions of the deformed gauge algebra, as well as the deformed reducibility
coefficients if the gauge transformations are reducible. The
problem is further complicated by the fact that one has to factor 
out the
``trivial" interactions that are simply induced by a change of variables. 

As we now review,
one can reformulate the problem as a cohomological problem
\cite{HenneauxBarnich93C}.  This
approach systematizes the recursive construction of the consistent
interactions and, furthermore, enables one to use the
powerful tools of homological algebra. 

Starting with a ``free" action
$\stackrel{(0)}{S_0} [\varphi^i]$ with ``free" gauge symmetries
\begin{equation}
\delta_\varepsilon \varphi^i = \stackrel{\smash{(0)}}{R}^i_\alpha
\varepsilon^\alpha ,
\end{equation}
leading to the Noether identities
\begin{equation}
{\delta\stackrel{(0)}{S}\over\delta\varphi^i}
\stackrel{\smash{(0)}}{R}^i_\alpha=0\ ,
\end{equation}
we introduce a coupling constant $g$ and modify $\stackrel{(0)}{S_0}$,
\begin{equation}
\stackrel{(0)}{S_0}\longrightarrow S_0 = \stackrel{(0)}{S_0} +
g \stackrel{(1)}{S_0} +
g^2 \stackrel{(2)}{S_0} + ...\label{fullaction}
\end{equation}
We consider only consistent deformations, meaning that the deformed action
should be gauge invariant as well. In the generic case this requires a
deformation of the gauge transformation rules,
\begin{equation}
\stackrel{\smash{(0)}}{R}^i_\alpha \longrightarrow R^i_\alpha =
\stackrel{\smash{(0)}}{R}^i_\alpha
+ g \stackrel{\smash{(1)}}{R}^i_\alpha +
g^2 \stackrel{\smash{(2)}}{R}^i_\alpha
+ ...\label{fullsymmetries} .
\end{equation}
Consistency is then translated into
the requirement that the Noether identities should hold to
all orders
\begin{equation}
{\delta{S}\over\delta\varphi^i}{R}^i_\alpha=0,\label{as1}
\end{equation}
where,
\begin{equation}
\delta_\varepsilon \varphi^i = R^i_\alpha \varepsilon^\alpha.
\end{equation}
Expanding Eq. (\ref{as1}) order by order in the coupling constant gives
consistency condition of increasing complexity.

For reducible theories, which is the case relevant to chiral $2$-forms,
there is an additional constraint.  The gauge transformations
of the free theory are not independent,
\begin{equation}
\stackrel{\smash{(0)}}{R}^i_\alpha \stackrel{\smash{(0)}}{Z}^\alpha_A
= 0
\end{equation}
(possibly on-shell).  One must then
also impose that the gauge transformations remain
reducible, possibly in a deformed way.  This yields additional
conditions on the coefficients $R$'s in Eq. (\ref{fullsymmetries}).

The deformations of an action fall into three classes. In the first one, gauge
invariant
terms are added to the original lagrangian and therefore no modification of the
gauge
transformations is required. Examples of this are functionals of the field
strength and its derivatives, as well as
Chern-Simons-like terms \cite{Deser:1982wh}. In the second
class, both the action and the transformation rules are modified.
However, the terms added to the transformation rules are invariant under the
original gauge transformations. As a consequence, the gauge algebra is not
modified to first order in the coupling constant. 
An example of this is the Freedman-Townsend model \cite{FT} for
two-forms in four dimensions. Finally, in the last class, the additional
terms in the deformed transformation rules are not gauge invariant.
Therefore the gauge algebra itself gets modified as well. The best known
example of this is the deformation of an abelian Yang-Mills theory to a
non-abelian theory.

The key to translating the problem of consistent interactions
into a cohomological problem is the antifield formalism
\cite{Batalin:1981jr,Batalin:1983wj,Batalin:1983jr}
(for reviews, see \cite{HenneauxTeitelboim92Book,Gomis:1995he}). 
Let us
assume that we solved the master equation for the undeformed theory. Its
solution is denoted by $\stackrel{(0)}{S}$, which satisfies
$(\stackrel{(0)}{S},\stackrel{(0)}{S})=0$. The existence of a consistent
deformation of the original gauge invariant action
implies the existence of a deformation of $\stackrel{(0)}{S}$,
which we denote by $S$,
\begin{equation}
\stackrel{(0)}{S}\longrightarrow S = \stackrel{(0)}{S} +
g \stackrel{(1)}{S} +
g^2 \stackrel{(2)}{S} + ...\label{fullmasterequation}
\end{equation}
Expanding the master equation for $S$, $(S,S)=0$, order by order in the
coupling constant yields various consistency relations,
\begin{eqnarray}
(\stackrel{(0)}{S},\stackrel{(0)}{S})&= 0 \label{deformation1}\\
(\stackrel{(0)}{S},\stackrel{(1)}{S})&= 0 \label{deformation2}\\
2(\stackrel{(0)}{S},\stackrel{(2)}{S}) +
(\stackrel{(1)}{S},\stackrel{(1)}{S})&= 0 \label{deformation3}\\
&\vdots\ \ \ .\nonumber
\end{eqnarray}
The first equation is satisfied by assumption. As $(\stackrel{(0)}{S},
(\stackrel{(0)}{S}, \cdot ))=0$, the second equation implies that
$\stackrel{(1)}{S}$
is a cocycle for the free differential $\stackrel{(0)}{s}\equiv
(\stackrel{(0)}{S},\cdot)$.
If $\stackrel{(1)}{S}$ is a coboundary, $\stackrel{(1)}{S}=
(\stackrel{(1)}{T},\stackrel{(0)}{S})$, one can show that this corresponds to a
trivial
deformation (i.e. a deformation which amounts to a simple redefinition of
the fields).

In practice, we consider deformations which are local in
spacetime, i.e., we impose
that $\stackrel{(1)}{S}, \stackrel{(2)}{S}, ...$ be {\em{local}}
functionals. Reformulating the equations in terms of the
Lagrange densities takes care of this problem.
E.g., rewriting equation (\ref{deformation2}) as
\begin{equation}
\stackrel{(0)}{s}\stackrel{(1)}{S}=\stackrel{(0)}{s}(\int
\stackrel{(1)}{\cal{S}})=0
\Leftrightarrow
\int(\stackrel{(0)}{s}\stackrel{(1)}{\cal{S}})=0,
\end{equation}
we obtain the following condition on the Lagrange density
$\stackrel{(1)}{\cal{S}}$,
\begin{equation}
\stackrel{(0)}{s}\stackrel{(1)}{\cal{S}}+d{\cal{M}}=0,
\label{blip}
\end{equation}
where ${\cal{M}}$ is a local form of degree $n-1$, where $n$ is the
dimensionality of
space-time and $d$ is the spacetime exterior
derivative\footnote{Throughout this paper we ignore boundary contributions}.
Again one can show that BRST-exact terms modulo $d$ are trivial solutions of
(\ref{blip}) and corresponds to trivial deformations.
In the local context, the proper cohomology to evaluate is thus
$H^{0,n}(\stackrel{(0)}{s}\mid d)$
where the first and second superscripts
denote the ghost number and form degree,
respectively.

Note that when all the representatives
of $H^{0,n}(\stackrel{(0)}{s}\mid d)$ can be taken not to depend
on the antifields, one may take
the first-order deformations
$\stackrel{(1)}{\cal{S}}$ to be antifield-independent.
In this case Eq. (\ref{deformation2}) reduces to
$(\stackrel{(0)}{S},\stackrel{(2)}{S})=0$ and implies that
the deformation at order $g^2$ defines also an element of
$H^{0,n}(\stackrel{(0)}{s}\mid d)$.  One can thus
take $\stackrel{(2)}{S}$ not to depend on the antifields either.  
Proceeding in this manner order by order in the coupling constant,
we conclude that the additional terms in $S$ are all
independent of the antifields.  Since the antifield-dependent
terms in the deformation of the master equation are related to
the deformations of the gauge transformations,
this means that there is no deformation of the gauge transformations.
Summarizing, if there is no non-trivial dependence on the antifields
in $H^{0,n}(\stackrel{(0)}{s}\mid d)=0$,
the only possible consistent interactions are of the first
class and do not modify the gauge symmetry.
This is the situation met for a system of
chiral $2$-forms, as we now pass to discuss.

\section{System of free chiral 2-forms in 6 dimensions}
\label{action}
\setcounter{equation}{0}
\setcounter{theorem}{0}
\setcounter{lemma}{0}

The non-covariant action for a system of N free chiral 2-forms is
\cite{HenneauxTeitelboim87ProcB},
\begin{equation}
S_0[A^A_{ij}]=\sum_A\int dtd^5x B^{Aij}(\dot{A}^A_{ij}-B^A_{ij}),\quad
(A=1,\ldots,N) ,
\label{flipflap}
\end{equation}
where
\begin{equation}
B^{Aij}=\frac{1}{6}\epsilon^{ijklm}F^A_{klm}=\frac{1}{2}\epsilon^{ijklm}
\partial_k A^A_{lm}.
\end{equation}
The integer $N$ can be any function of the number $n$ of coincident M5-branes
(e.g., $N \sim n^3$).  
The action (\ref{flipflap})
differs from the one in \cite{PS}-\cite{APPS} where a
space-like dimension was singled out. Here we take time as the
distinguished direction; from the point of view of the
PST formulation \cite{PST,BLNPST}, 
the two approaches simply differ in the gauge fixation.
We work in Minkowski spacetime.  This implies, in particular, that
the topology of the spatial sections $R^5$ is trivial.  
Most of our considerations would go unchanged in a curved background
of the product form $R \times \Sigma$
provided the De Rham cohomology groups $H^2_{DeRham}(\Sigma)$ and
$H^1_{DeRham}(\Sigma)$ of the spatial sections $\Sigma$ vanish.
[If $H^2_{DeRham}(\Sigma)$ is non-trivial, there are additional
gauge symmetries besides (\ref{cling}) below, given by time-dependent
spatially closed $2$-forms; 
similarly, if $H^1_{DeRham}(\Sigma)$ is non-trivial,
there are additional reducibility identities besides (\ref{clong})
below.  One would thus need additional
ghosts and ghosts of ghosts.
These, however, would not change the discussion of {\em local}
Lagrangians because they would be global in space (and local in $t$).]

The action $S_0$ is invariant under the following gauge transformations
\begin{equation}
\delta_{\Lambda}A^A_{ij}=\partial_i \Lambda^A_j - \partial_j \Lambda^A_i,
\label{cling}\end{equation}
because $B^{Aij}$ is gauge-invariant and identically transverse ($\partial _i
B^{Aij} \equiv 0$)
\footnote{%
Since $A^A_{0i}$ does not occur in the action -- even if one replaces $%
\partial_0 A^A_{ij}$ by $\partial_0 A^A_{ij} - \partial_i A^A_{0j} - \partial_j
A^A_{i0}$ (it drops out because $B^{Aij}$ is transverse) --, the action is of
course invariant under arbitrary shifts of $A^A_{0i}$.}.
As $\delta A_{ij}^A=0$ for 
\begin{equation}
\Lambda_i^A=\partial_i\varepsilon^A,
\label{clong}
\end{equation}
this set of gauge transformations is reducible.
This exhausts completely the redundancy in $\Lambda_i^A$
since $H^1_{DeRham}(R^5) = 0$.

The equations of motion obtained from $S_0[A^A_{ij}]$ by varying $A^A_{ij}$ are
\begin{equation}
\epsilon^{ijklm}\partial_k \dot{A}^A_{lm}-2\partial_k F^{Aijk}=0
\Leftrightarrow
\epsilon^{ijklm}\partial_k (\dot{A}^A_{lm}-B^A_{lm})=0.
\label{bam}
\end{equation}
Using $H^2_{DeRham}(R^6)=0$,  one finds that
the general solution of (\ref{bam}) is
\begin{equation}
\dot{A^A_{ij}}-B^A_{ij}=\partial_i \Lambda^A_j - \partial_j \Lambda^A_i.
\label{solution}
\end{equation}
The ambiguity in the solutions of the equations of motion is thus completely
accounted for by the gauge freedom (\ref{cling}).
Hence the set of gauge transformations is complete.

We can view $\Lambda^A_i$ as $A^A_{0i}$, so the equation (\ref{solution}) can be
read as the self-duality equation
\begin{equation}
F^A_{0ij}-*F^A_{0ij}=0,
\end{equation}
where $F^A_{0ij}=\dot{A^A_{ij}}+\partial_i A^A_{j0}+\partial_j A^A_{0i}$.
Alternatively, one may use the
gauge freedom to set $\Lambda^A_i=0$, which yields the self-duality condition
in the
temporal gauge.

\section{Fields - Antifields - Solution of the master equation}
\setcounter{equation}{0}
\setcounter{theorem}{0}
\setcounter{lemma}{0}

The solution of the master equation is easy to construct in this
case because the gauge transformations are abelian.
We refer to \cite{Batalin:1981jr,Batalin:1983wj,Batalin:1983jr,%
HenneauxTeitelboim92Book} for the general construction.

The fields in presence here are
\begin{equation}
\{\Phi^M\}=\{A^A_{ij},C^A_i,\eta^A\}.
\end{equation}
The ghosts $C^A_i$ corresponds to the gauge parameters $\Lambda^A_i$,
and the ghosts of ghosts $\eta^A$ corresponds to $\epsilon^A$.

Now, to each field $\Phi^M$ we associate an antifield $\Phi^*_M$. The set of
antifields is then
\begin{equation}
\{\Phi^*_M\}=\{A^{*Aij},C^{*Ai},\eta^{*A}\}.
\end{equation}
The fields and antifields have the respective parities
\begin{eqnarray}
&&\epsilon(A^A_{ij})=\epsilon(\eta^A)=\epsilon(C^{*Ai})=0\\
&&\epsilon(C^A_i)=\epsilon(A^{*Aij})=\epsilon(\eta^{*A})=1.
\end{eqnarray}
The antibracket is defined as
\begin{equation}
(X,Y)=\int d^nx\left( \frac{\delta^RX}{\delta\Phi^M(x)}
\frac{\delta^LY}{\delta\Phi^*_M(x)}
-\frac{\delta^RX}{\delta\Phi^*_M(x)}\frac{\delta^LY}{\delta\Phi^M(x)}\right)
\end{equation}
where $\delta^R/\delta Z(x)$ and $\delta^L/\delta Z(x)$
denote functional right- and left-derivatives.

Because the set of gauge transformations is complete and defines a closed
algebra,
the (minimal, proper) solution of the master equation $(S,S)=0$
takes the general form
\begin{equation}
S=S_0+\sum_M\int (-)^{{\epsilon}(M)}\Phi^*_M s\Phi^M,
\end{equation}
where ${{\epsilon}(M)}$ is the Grassmann parity of $\Phi^M$. 
More explicitly, we have
\begin{equation}
S=S_0 +\sum_A \int dtd^5x(A^{*Aij}\partial_i C^A_j - C^{*Ai}\partial_i \eta^A)
\end{equation}

The solution $S$ of the master equation captures all the information about
the gauge structure of the theory : the Noether identities, the 
closure of the gauge transformations and the higher order gauge identities are
contained in the master equation. The existence of $S$ reflects the consistency
of the gauge
transformations.

\section{BRST operator}
\setcounter{equation}{0}
\setcounter{theorem}{0}
\setcounter{lemma}{0}

The BRST operator $s$ is obtained by taking the antibracket with
the proper solution $S$ of the classical master equation,
\begin{equation}
s\, X   = (S,X).
\end{equation}

The BRST operator can be decomposed as
\begin{equation}
s=\delta+\gamma
\end{equation}
where $\delta$ is the Koszul--Tate differential \cite{HenneauxTeitelboim92Book}.
What distinguishes $\delta$ and $\gamma$ is the antighost number ($antigh$)
defined through
\begin{eqnarray}
antigh(A^A_{ij})=antigh(C^A_i)=antigh(\eta^A)=0,\\
antigh(A^{*Aij})=1,\quad antigh(C^{*Ai})=2,\quad antigh(\eta^{*A})=3.
\end{eqnarray}

The ghost number ($gh$) is related to the antighost number by
\begin{equation}
gh=puregh-antigh
\end{equation}
where $puregh$ is defined through
\begin{eqnarray}
puregh(A^A_{ij})=0, \quad puregh(C^A_i)=1,\quad puregh(\eta^A)=2,\\
puregh(A^{*Aij})=puregh(C^{*Ai})=puregh(\eta^{*A})=0.
\end{eqnarray}

The differential $\delta$ is characterized by $antigh(\delta)=-1$, i.e. it
lowers the antighost number by one unit and acts on the fields and antifields
according to
\begin{eqnarray}
\delta A^A_{ij}&=&\delta C^A_{i}=\delta \eta^A=0,\\
\delta A^{*Aij}&=&2\partial_k F^{Akij}-\epsilon^{ijklm}\partial_k
\dot{A}^A_{lm},\\
\delta C^{*Ai}&=&\partial_j A^{*Aij},\\
\delta \eta^{*A} &=& \partial_i C^{*Ai}.
\end{eqnarray}

The differential $\gamma$ is characterized by $antigh(\gamma)= 0$ and acts as
\begin{eqnarray}
\gamma A^A_{ij}&=&\partial_i C^A_j - \partial_j C^A_i,\\
\gamma C^A_i&=&\partial_i \eta^A,\label{yep}\\
\gamma \eta^A&=&0,\\
\gamma A^{*Aij}&=&\gamma C^{*Ai}=\gamma \eta^{*A}=0.
\end{eqnarray}

Furthermore we have,
\begin{equation}
sx^\mu = 0, \; s(dx^\mu)=0.
\end{equation}

\section{Local forms - Algebraic Poincar\'e lemma}
\setcounter{equation}{0}
\setcounter{theorem}{0}
\setcounter{lemma}{0}

A $\em{local}$ $\em{function}$ is a function 
of the fields,
the ghosts, the antifields,  and their derivatives up to some finite
order $k$ (which depends on the function),
\begin{equation}
f=f(\Phi,\partial_{\mu}\Phi,\ldots,\partial_{\mu_1}\ldots\partial_{\mu_k}\Phi).
\end{equation}
A \emph{local function} is thus a function over a finite dimensional vector
space $J^k$ called  ``jet space".
A \emph{local form} is an exterior polynomial in the $dx^\mu$'s
with local functions as
coefficients.
The algebra of local forms will be denoted by
${\cal{A}}$.  In practice, the local forms are polynomial in
the ghosts and the antifields, as well as in the differentiated
fields, so we shall from now on assume that the local forms under
consideration are of this type. One can actually show that polynomiality
in the ghosts, the antifields and their derivatives follows from
polynomiality in the derivatives of the $A_{ij}$ by an argument similar
to the one used in \cite{BBH2} for $1$-forms; and polynomiality
in the derivatives is automatic in our perturbative approach
where we work order by order in the coupling constant(s).

Note also that we exclude an explicit $x$-dependence
of the local forms.  One could allow for one without change
in the conclusions.  In fact, as we shall indicate
below, allowing for an explicit $x$-dependence simplifies 
some of the proofs.  We choose not to do so here since the interaction
terms in the Lagrangian should not depend explicitly on the
coordinates in
the Poincar\'e-invariant context.

The following
theorem describes the cohomology of $d$ in the algebra of local forms,
in degree $q<n$. 
\begin{theorem}
The cohomology of $d$ in the algebra of local forms of degree $q<n$ is given
by
\begin{eqnarray}
H^0(d) &\simeq& R, \nonumber\\
H^q(d)&=& \{ \hbox{Constant Forms} \} , \; 0<q<n .\nonumber
\end{eqnarray}
\end{theorem}
Constant forms are by definition polynomials in the $dx^\mu$'s
with constant coefficients.
This theorem is called the algebraic Poincar\'e lemma (for $q<n$).
There exist many proofs of this lemma
in the literature.  One of the earliest
can be found in \cite{Vinogradov77,Vinogradov78}.

Constant $q$-forms are trivial in degree $0<q<n$ in the algebra of
local forms with an explicit $x$-dependence; e.g., $dx^0 = df$,
where $f$ is the $x^0$-dependent function $f= x^0$.  Thus, in this
enlarged algebra, the cohomology of $d$ is simpler and vanishes
in degrees $0<q<n$.  This is the reason that the calculations are
somewhat simpler when one allows for an expicit $x$-dependence.

We work in a formalism where the time direction is privileged. For this
reason,
it is useful to introduce the following notation : the $l$-th time
derivative of
a field $\Phi$ (including the ghosts and antifields) is denoted by
$\Phi^{(l)}$ ($=\partial_0^l\Phi$), and
the spatial differential is denoted by $\tilde{d}=dx^i\partial_i$.

A \emph{local spatial form} is an exterior polynomial in the
spatial $dx^k$'s with coefficients that are local functions.
If we write the set of the generators of the jet space $J^k$ as
\begin{equation}
\{\Phi^{(l_0)},\partial_{i_1}\Phi^{(l_1)},\ldots,
\partial_{i_1}\dots\partial_{i_k}\Phi^{(0)};\,\,l_j=0,\ldots,k-j\},
\end{equation}
it is clear that
\begin{theorem}
The cohomology
of $\tilde{d}$ in the algebra of local spatial forms of degree $q<n-1$ is given
by
\begin{eqnarray}
H^0(\tilde{d}) &\simeq& R, \nonumber\\
H^q(\tilde{d}) &=& \{ \hbox{Constant spatial  forms} \},
\; 0<q<n-1 .\nonumber
\end{eqnarray}
\end{theorem}

A similar decomposition of space and time derivatives occurs of course
in the Hamiltonian formalism. A discussion of the problem of
consistent deformations of a gauge invariant action has
been carried out in the Hamiltonian context
in \cite{Bizdadea:2000ha,Bizdadea:2000hb,Bizdadea:2000hi}.
 
\section{Cohomology of $\gamma$}
\setcounter{equation}{0}
\setcounter{theorem}{0}
\setcounter{lemma}{0}

The following theorem completely gives $H(\gamma)$.
\begin{theorem}
The cohomology of $\gamma$ is given by,
\begin{equation}\label{gamma}
H(\gamma)={\cal{I}}\otimes V.
\end{equation}
Here, the algebra ${\cal{I}}$ is the algebra of
the local forms with coefficients that depend only on the variables 
$F^A_{ijk}$, the antifields $\phi^*_M$,
and all their partial derivatives up to a finite order
(``gauge-invariant" local forms). These variables are collectively
denoted by $\chi$.
The algebra $V$ is the polynomial algebra in the
ghosts $\eta^A$ of ghost number two and their time derivatives.
\end{theorem}
\proof{
The generators of ${\cal{A}}$ can be grouped in three sets:
\begin{eqnarray}
&&T=\{t^i\}=\{\partial^{}_{{\mu}_1 \ldots
{\mu}_k}F^{A}_{ijk},\partial^{}_{{\mu}_1 \ldots
{\mu}_k}\phi_M^*,\eta^{A(l)},dx^{\mu}\}\\
&&U=\{u^{\alpha}\}=\{\partial^{}_{(i_1 \ldots i_k}A^{A(l)}_{[i)_2
j]_1},\partial^{}_{(i_1 \ldots i_{k-1}}C^{A(l)}_{i_k)}\}\\
&&V=\{v^{\alpha}\}=\{\partial^{}_{i_1 \ldots
i_k}\partial^{}_{[i}C^{A(l)}_{j]},\partial^{}_{i_1 \ldots i_k}\eta^{A(l)}\}
\end{eqnarray}
($k,l = 0, \cdots$)
where $[$ $]$ and $($ $)$ mean respectively antisymmetrization and
symmetrization; the subscript indicates
the order in which the operations are made.

The differential $\gamma$ acts on these three sets in the following way
\begin{equation}
\gamma T=0,\quad \gamma U=V,\quad \gamma V=0.
\end{equation}
The elements of $U$ and $V$ are in a one-to-one correspondence
and are linearly independent with respect to each other,
so they
constitute a manifestly
contractible part of the algebra and can thus be removed from the cohomology.

No element in the algebra of generated by  $T$ is trivial in the cohomology
of $\gamma$, except $0$.
Indeed, let us assume the existence of a local form $F(t^i)\neq 0$ which is
$\gamma$-exact, then
\begin{eqnarray}
F(t^i)=\gamma G(t^i,u^{\alpha},v^{\alpha})=v^{\alpha}\frac{\partial^L
G}{\partial u^{\alpha}}(t^i,u^{\alpha},v^{\alpha}).
\end{eqnarray}
But this implies that
\begin{equation}
F(t^i)=F(t^i)\mid_{v^{\alpha}=0}=0,
\end{equation}
as announced.
}

Note that contrary to what happens in the non-chiral case, the
temporal derivatives of the ghosts $\eta^A$ are non-trivial in
cohomology.  There is thus an infinite number of generators
in ghost number two for $H(\gamma)$, namely, all the
$\eta^{A(l)}$'s.  In contrast, in the non-chiral case, one
has $\partial_0\eta^A = \gamma C^A_0$ and so
$\partial_0\eta^A$ (and all the subsequent derivatives)
are $\gamma$-exact.  In the chiral case, there is no $C^A_0$.

Let$\{\omega^I\}$ be a basis of the vector space $V$ of polynomials
in the variables $\eta^A$ and all their
time derivatives.
Theorem \ref{gamma} tells us that
\begin{equation}
\gamma \alpha =0, \;
\alpha \in \cal{A} \quad\Leftrightarrow \quad\alpha =\sum_I P_I(\chi)\omega^I+
\gamma\beta.
\end{equation}
Furthermore, because $\omega^I$ is a basis of $V$
\begin{equation}
\sum_I P_I(\chi)\omega^I=\gamma\beta \quad\Rightarrow \quad P_I(\chi)=0.
\end{equation}

It will be useful in the sequel to 
choose a special basis $\{ \omega^I \}$. The vector space $V$ of
polynomials in the ghosts $\eta^A$ and their time derivatives
splits as the direct sum $V^{2k}$ of vector spaces with definite
pure ghost number $2k$.  The space $V^0$ is one-dimensional and 
given by the constants. We may choose $1$ as basis vector for $V^0$,
so let us turn to the less trivial spaces $V^{2k}$ with $k \not= 0$.
These spaces are themselves the direct sums of 
finite dimensional vector spaces $V^{2k}_r$ containing
the 
polynomials with exactly $r$ time derivatives of the $\eta$'s
(e.g., $\partial_0 \eta^A \, \partial_{00} \eta^B$ is in
$V^{4}_3$).  
The following lemma provides a basis of $V^{2k}$ for $k \not=0$:
\begin{lemma}
Let $V^{2k}$ be the vector space of polynomials in the 
variables $\eta^{A(l)}$
with fixed pure ghost number $2 k \neq 0$.
$V^{2k}$ is the direct sum
\begin{equation}
V^{2k}=V^{2k}_0\oplus V^{2k}_1 \oplus
\ldots,
\end{equation}
where $V^{2k}_m$ is the subspace of $V^{2k}$ 
containing the polynomials with exactly
$m$ derivatives of $\eta^A$.  One has
dim$V^{2k}_m\leq$ dim$V^{2k}_{m+1}$.
There exist a basis of $V^{2k}_m$
\begin{equation}
\{\omega^{I_m}_{(m)}:\,I_m=1,\ldots,q_m;\, m=0,\ldots\},
\end{equation}
which fulfills
\begin{equation}
\omega^{I_m}_{(m)}=\partial_0\omega^{I_m}_{(m-1)}\quad (I_m=1,\ldots,q_{m-1}).
\label{ploc}
\end{equation}
In other words, the first $q_{m-1}$ basis vectors of $V^{2k}_m$ are directly
constructed from the basis vectors of $V^{2k}_{m-1}$
by taking their time derivative $\partial_0$.
\label{zap}
\end{lemma}
\proof{We will prove the lemma by induction. For $m=0$,
take an arbitrary basis of $V^{2k}_0$ 
(space of polynomials in the undifferentiated
ghosts $\eta^A$ of degree $k$). Assume now that a basis with
the required properties exists up to order $m-1$.
Let $\{\omega^I_{(m-1)};\, I=0,\ldots,q_{m-1}\}$ be
a basis with those properties for $V^{2k}_{m-1}$.
We want to prove that it is possible to construct a basis of $V^{2k}_m$
where the first $q_{m-1}$ basis vectors are the time derivatives of the
basis vectors of $V^{2k}_{m-1}$.
We only have to show that the $\partial_0\omega^I_{(m-1)}$ are linearly
independent (because they can always
be completed to form a basis of $V^{2k}_{m}$). In other words, we must prove that
\begin{equation}
\sum \limits_{I=1}^{q_{m-1}}\lambda_I\partial_0\omega^I_{(m-1)}
=\partial_0(\sum \limits_{I=1}^{q_{m-1}}\lambda_I\omega^I_{(m-1)})=0
\label{bardaf}
\end{equation}
implies $\lambda_I=0$. But (\ref{bardaf}) is equivalent to
\begin{equation}
\sum \limits_{I=1}^{q_{m-1}}\lambda_I\omega^I_{(m-1)}=K,
\end{equation}
where $K$ is a constant (algebraic Poincar\'e lemma in form
degree 0). $K$ must be equal to zero
because we are in pure ghost number $\neq 0$.
By hypothesis, the $\omega^I_{(m-1)}$ are linearly independant, hence
the $\lambda_I$ must be all equal to zero, which ends the proof.
}

\section{Cohomology of $\gamma$ modulo $d$ at positive antighost number}
\label{Hgammad}
\setcounter{equation}{0}
\setcounter{theorem}{0}
\setcounter{lemma}{0}

Let be $a^p$ a local $p$-form of antighost number $k\neq 0$ fulfilling
\begin{equation}
\gamma a^p + db^{p-1}=0.
\label{gammad}
\end{equation}
We want to show that if we add to $a^p$ an adequate $d$-trivial term, the
equation
(\ref{gammad}) reduces to $\gamma a^p=0$.

{}From (\ref{gammad}), using the algebraic Poincar\'e lemma and the fact that
$\gamma$
is nilpotent and anticommute with $d$, we can derive the descent equations
\begin{eqnarray}
\gamma a^p &+& db^{p-1}=0\\
\gamma b^{p-1} &+&dc^{p-2}=0\\
&\vdots&\nonumber\\
\gamma e^{q+1}&+&df^q=0 \label{before}\\
\gamma f^{q}&=&0\label{last},
\end{eqnarray}
Indeed, the fact that the antighost number is strictly positive eliminates the
constants. [E.g., from (\ref{gammad}), one derives $d \gamma b^{p-1} = 0$
and thus $\gamma b^{p-1} +dc^{p-2}= \hbox{constant}$, but the
constant must vanish since it must have strictly positive
antighost number.]
We suppose $q<p$, since otherwise $\gamma a^p =0$,
which is the result we want to prove.
The equation (\ref{last}) tells us that $f^q$ is a cocycle of $\gamma$.
It must be non-trivial in 
$H^q(\gamma)$ because if $f^q =
\gamma g^q$,
then (\ref{before}) becomes $\gamma (e^{q+1}-dg^q)=0$.
The redefinition $e^{'q+1}=e^{q+1}-dg^q$ does not affect the descent equation
before (\ref{before}), which means that the descent
stops one step earlier, at $q-1$.

Using theorem \ref{gamma}, we deduce from (\ref{last}) that
\begin{equation}
f^q =\sum_{m,I_m}[\tilde{P}^{(m)}_{I_m}(\chi)
+ dx^0\tilde{Q}^{(m)}_{I_m}(\chi)]\omega^{I_m}_{(m)},
\label{bzz}
\end{equation}
where $\tilde{P}^{(m)}_{I_m}$ and $\tilde{Q}^{(m)}_{I_m}$
are local spatial forms of respective degree $q$ and $q-1$. We take the
basis elements $\omega^{I_m}_{(m)}$ to fulfill the
conditions of lemma \ref{zap}.
Differentiating (\ref{bzz}), we find
\begin{eqnarray}
df^q&=&\sum_{m,I_m}\{\tilde{d}\tilde{P}^{(m)}_{I_m}\omega^{I_m}_{(m)}
+\gamma(\tilde{P}^{(m)}_{I_m}\hat{\omega}^{I_m}_{(m)})\nonumber \\
&&+dx^0[(\partial_0\tilde{P}_{I_m}^{(m)}-
\tilde{d}\tilde{Q}_{I_m}^{(m)})\omega^{I_m}_{(m)}
+\tilde{P}^{(m)}_{I_m}\partial_0\omega^{I_m}_{(m)}]\}.
\label{afterbefore}
\end{eqnarray}
The local function $\hat{\omega}^{I_m}_{(m)}$ is defined by
$\tilde{d}\omega^{I_m}_{(m)}=\gamma\hat{\omega}^{I_m}_{(m)}$ (
and exists thanks to equation
(\ref{yep})).

Now, we will show that the component $\tilde{P}^{(m)}_{I_m}$ can be eliminated
from $f^q$ by a trivial redefinition of $f^q$.
In order to satisfy (\ref{before}), the term 
independent of $dx^0$ and the coefficient of the term linear
in $dx^0$ in (\ref{afterbefore}) must separately
be $\gamma$-exact. The second condition gives explicitly 
\begin{equation}
\sum_{m,I_m}[(\partial_0\tilde{P}_{I_m}^{(m)}-\tilde{d}
\tilde{Q}_{I_m}^{(m)})\omega^{I_m}_{(m)}
+\tilde{P}^{(m)}_{I_m}\partial_0\omega^{I_m}_{(m)}]=\gamma\beta,
\label{back2}
\end{equation}
To analyze precisely this equation, we define a degree $T$ by
\begin{equation}
T(\chi)=0,\quad T(\eta^{A(m)})=m.
\end{equation}
In fact, $T$ simply counts the number of time derivative of $\eta^A$.
We can decompose (\ref{back2}) according to the degree $T$.
Let $p$ be the highest degree occuring
in $f^q$.  Then, the highest degree occuring in
(\ref{back2}) is $p+1$  and we must have
\begin{equation}
\sum_{I=1}^{q_p}\tilde{P}^{(p)}_{I}
\partial_0\omega^{I}_{(p)}=\gamma\beta_{p+1}.
\end{equation}
{ } From the proof of the lemma \ref{zap}, we find that
\begin{eqnarray}
\tilde{P}^{(p)}_I=0\quad (I=1,\ldots,q_p)
\end{eqnarray}
because the $\partial_0\omega^{I}_{(p)}$ are linearly independent.
In $T$-degree $p$, (\ref{back2}) gives then
\begin{eqnarray}
\gamma\beta_p&=&-\sum_{I=1}^{q_p}\tilde{d}\tilde{Q}_{I}^{(p)}\omega^{I}_{(p)}
+\sum_{I=1}^{q_{p-1}}\tilde{P}_I^{(p-1)}\partial_0\omega^I_{(p-1)}\\
&=&\sum_{I=1}^{q_{p-1}}(\tilde{P}_I^{(p-1)}\,-
\tilde{d}\tilde{Q}_I^{(p)})\omega^I_{(p)}
-\sum_{I=q_{p-1}+1}^{q_p}\tilde{d}\tilde{Q}_I^{(p)}\omega^I_{(p)},
\end{eqnarray}
where we have used the property (\ref{ploc}) of the basis $\{\omega^I \}$.
This implies that
\begin{equation}
\tilde{P}_I^{(p-1)}=\tilde{d}\tilde{Q}_I^{(p)}\quad (I=1,\ldots,q_{p-1})
\end{equation}
Inserting this equation in (\ref{bzz}), we find that $\tilde{P}_I^{(p-1)}$ can
be removed from $f^q$ by eliminating a trivial cocycle of $\gamma$ modulo $d$
and redefining $\tilde{Q}_I^{(p-1)}$. It only affects $e^{q+1}$ by a 
$d$-exact term.
Next, the equation (\ref{back2}) at $T$-degree $p-1$ shows that
$\tilde{P}_I^{(p-2)}$
is also $\tilde{d}$-exact and can thus also be removed.
Proceeding in the same way until the order 1 in $T$, we have proved that
all the $\tilde{P}^{(m)}_I$ can be eliminated from $f^q$.

Looking back at (\ref{back2}) and taking into account that
$\tilde{P}^{(m)}_{I_m}$ can be set equal to zero by the
above argument, we find that
\begin{equation}
\tilde{d}\tilde{Q}^{(m)}_{I_m}=0.
\label{tilt}
\end{equation}
Now, we must use the invariant Poincar\'e lemma
(invariant means in the algebra ${\cal{I}}$ of
gauge-invariant forms) stating that
\begin{theorem}\label{invP}
Let be $\tilde{P}(\chi)$ a local spatial form of degree $q<5$, then
\begin{equation}
\tilde{d}\tilde{P}(\chi)=0\Rightarrow
\tilde{P}(\chi)=\tilde{R}(F^{A(l)})+\tilde{d}\tilde{Q}(\chi),
\label{blub}
\end{equation}
where $\tilde{R}(F^{A(l)})$ is a polynomial in the curvature forms
$F^A=\frac{1}{6}F^A_{ijk}dx^i dx^j dx^k$
and all their time derivatives (with coefficients that may
involve $dx^k$, which takes care of the constant forms).
\end{theorem}
\proof{The set of the generators of the algebra ${\cal{I}}$ is
\begin{equation}
\{\chi\}=\{\partial^{}_{i_1 \ldots i_k}F^{A(l)}_{ijk},\partial^{}_{i_1 \ldots
i_k}\phi^{*(l)}_M,\eta^{A(l)},dx^{\mu}\}
\label{setsetset}
\end{equation}
The 1-form $dx^0$ is not present in our problem since $\tilde{P}$
is a spatial local form (it only involves $dx^k$).
Considering $l$ and $A$ as only one label (call
it $\alpha$) and forgetting about $dx^0$,
the set (\ref{setsetset}) is the same as the corresponding set
of generators of the algebra ${\cal{I}}$($\equiv H(\gamma)$ 
in pureghost number
0) for a system of spatial two-forms $ \{A^\alpha_{ij} \equiv
A^A_{ij}, \partial_0 A^A_{ij}, \partial_{00}A^A_{ij}, \cdots \}$
in 5 dimensions.
Consequently, we can simply use the results demonstrated in
\cite{HenneauxKnaepenSchomblond96}
for a system of $p$-forms in any dimension.}
We assumed before that $f^q$ is of degree $q<6$, hence $\tilde{Q}_I$ is of
degree $<5$. Thus, (\ref{tilt}) implies
\begin{equation}
\tilde{Q}^{(m)}_{I_m}=\tilde{d}\tilde{R}^{(m)}_{I_m},
\label{baff}
\end{equation}
where $\tilde{R}^{(m)}_{I_m}$ is a spatial form which only depends on the
variables $\chi$. There is no exterior polynomial in the curvatures
in $\tilde{Q}^{(m)}_{I_m}$ because $\tilde{Q}^{(m)}_{I_m}$ has
strictly positive antighost number.
We can therefore conclude that $f^q$ is trivial in $H^q(\gamma\mid d)$ and can
be eliminated by redefining $e^{q+1}$.
The true bottom is then one step higher. We can proceed in the same way until
we arrive at $\gamma a^{'p}=0$ with $a^{'p}=a^{p}+dg^{p-1}$.
This can be translated into the following theorem
\begin{theorem}\label{gammad'}
Let be a local form $a$ of antighost number $\neq 0$ fulfilling
$\gamma a + db=0$. There exists a local form $c$ such as $a':=a+dc$ satisfies
$\gamma a'=0$.
\end{theorem}

\section{Cohomology of $\gamma$ modulo $d$ at zero antighost number}
\label{Hgammadbis}
\setcounter{equation}{0}
\setcounter{theorem}{0}
\setcounter{lemma}{0}

Now, we want to study $H^{6,0}(\gamma\mid d)$ in pureghost number 0. Let be
$a^{(6,0)}\in{\cal{A}}$ of form degree 6, of antighost and pureghost number 0,
and fulfilling
$\gamma a^{(6,0)} + da^{(5,1)}=0$.
If $a^{(5,1)}$ is trivial $\gamma$ modulo $d$, this equation reduces 
to $\gamma
a^{(6,0)}+db^{(5,0)}=0$, which gives
$a^{(6,0)}=f(\partial^{}_{{\mu}_1 \ldots {\mu}_k}F^{A}_{ijk})d^6x$ plus
a term trivial in the cohomology of $\gamma$ modulo $d$.

Otherwise, we can derive the non trivial descent equations
\begin{eqnarray}
\gamma a^{(6,0)} &+& da^{(5,1)}=0\label{first}\\
\gamma a^{(5,1)} &+&da^{(4,2)}=0\\
&\vdots&\nonumber\\
\gamma a^{(7-g,g-1)}&+&da^{(6-g,g)}=0 \label{before'}\\
\gamma a^{(6-g,g)}&=&0\label{last'},
\end{eqnarray}
because $pureghost(\gamma a^{(6-i,i)})>0$ eliminates the constants.
If $a^{(6-g,g)}$ is trivial $\gamma$ modulo $d$, the bottom is really one step
higher.

Eq. (\ref{last'}) implies that 
\begin{equation}
a^{(6-g,g)}=\sum_I(\tilde{P}^{6-g}_I(\chi)+
dx^0\tilde{Q}^{5-g}_I(\chi))\omega^I+\gamma b^{(6-g,g-1)},
\label{last''}
\end{equation}
where $\tilde{P}^{6-g}_I$ and $\tilde{Q}^{5-g}_I$ are local spatial forms, the
superscript giving the form degree.
Because the pureghost number of $\eta$ is two, $a^{(6-g,g)}$ is non trivial
only for $g$ even.
So, three cases are of interest: $g=0,2,4$.

The case $g=0$ corresponds to
$\gamma a^{(6,0)} = 0$ and has been already studied so let
us assume $g>0$.
The equations (\ref{before'}) and (\ref{last''}) imply together
\begin{equation}
\sum_I(\partial_0
\tilde{P}^{6-g}_I-\tilde{d}\tilde{Q}^{5-g}_I)\omega^I+
\sum_I\tilde{P}^{6-g}_I\partial_0\omega^I=\gamma\beta.
\label{blob}
\end{equation}
Repeating the same analysis as for the equation (\ref{back2}), we arrive at
the conclusion that
$\tilde{P}^{6-g}_I$ is trivial in the invariant cohomology of $\tilde{d}$ (or
vanishes)
and can thus be removed from $a^{(6-g,g)}$ by the addition of trivial terms in
the cohomology of $\gamma$ modulo $d$
and a redefinition of $\tilde{Q}^{5-g}_I$.
The case $g=6$ is then eliminated because in that case $\tilde{Q}^{5-g}_I$ is
not present at all.
Hence, there remains only two cases to examine: $g=2$ and $g=4$.

Once $\tilde{P}^{6-g}_I$ is removed, the equation (\ref{blob}) gives
$\tilde{d}\tilde{Q}^{5-g}_I=0$.
Using the invariant Poincar\'e lemma, we find
$\tilde{Q}^{5-g}_I=\tilde{R}^{5-g}_I(F^{A(l)})+
\tilde{d}\tilde{S}^{(4-g,g)}_I(\chi)$.
Hence, the form of the bottom is
\begin{equation}
a^{(6-g,g)}=dx^0\sum_I\tilde{R}^{5-g}_I(F^{A(l)})\omega^I+\gamma
b^{(6-g,g-1)}+dc^{(5-g,g)}.
\end{equation}
But $F^{A(l)}$ is of form degree 3, thus if $g=4$, $\tilde{R}^{5-g}_I$
must be a constant spatial $1$-form.  In that instance, the $\omega^I$
must be quadratic in the ghosts $\eta^{A(l)}$. The lift of such a
bottom is obstructed (i.e., leads to no $a^{6,0}$) unless
it is trivial (see \cite{knaepen}), 
so that the case $g=4$ need not be considered. [In the algebra of
$x$-dependent local forms, the argument is simpler:
the bottom is always trivial and removable
since it involves a constant 1-form, which is trivial.] 

It only remains to examine the case $g=2$.
$\tilde{R}$ must then be a 3-form. One can take $\tilde{R}$ linear in $F^{A(l)}$.
In that case, the lift gives Chern-Simons terms, which are linear combinations
of $dx^0F^{A(l)}A^{B(m)}$,
with $A^{B(m)}=\frac{1}{2}A^{B(m)}_{ij}dx^idx^j$. Or one can take $\tilde{R}$
to be a constant 3-form. The corresponding deformation is linear in the 2-form $A^{A(l)}$
with coefficients that are constant forms. 
This second possibility is not $SO(5)$ invariant and leads to equations of motion that are not
Lorentz invariant. It will not be considered further.

Dropping the latter possibility, all these results can be summarized in the
\begin{theorem}\label{inter}
The non trivial elements of $H^{6,0}_0(\gamma\mid d)$ are of two types:
(i) those that descend trivially; they are of the form $f(\partial^{}_{{\mu}_1
\ldots {\mu}_k}B^{A}_{ij})d^6x$;
(ii) those that descend non trivially; they are linear combinations of the
Chern-Simons terms
$\partial^l_0B^{Aij}\partial^m_0A^B_{ij}d^6x$.
\end{theorem}

Note that the kinetic term in the free action is precisely of the
Chern-Simons type (with $l=0$ and $m=1$).

\section{Invariant cohomology of $\delta$ modulo $\tilde{d}$ in antighost
number $2,4,6,\ldots$}
\setcounter{equation}{0}
\setcounter{theorem}{0}
\setcounter{lemma}{0}

To pursue the analysis, we  need some results on
the cohomology of the Koszul-Tate differential $\delta$
as well as on its
mod-$d$ and mod-$\tilde{d}$ cohomologies.

We can rewrite the action of the Koszul-Tate differential in the following way
\begin{eqnarray}
\delta A^{A(l)}_{ij}&=&\delta C^{A(l)}_{i}=\delta \eta^{A(l)}=0,\\
\delta A^{*A(l)ij}&=&2\partial_k F^{A(l)kij}-\epsilon^{ijklm}\partial_k
A^{A(l+1)}_{lm},\\
\delta C^{*A(l)i}&=&\partial_j A^{*A(l)ij},\\
\delta \eta^{*A(l)} &=& \partial_i C^{*A(l)i}.
\end{eqnarray}
If we regard $A$ and $l$ as only one label, these equations corresponds to an
infinite
number of coupled non-chiral 2-forms in 5 dimensions.

It is useful to introduce a degree $N$ defined as
\begin{eqnarray}
&& N(\Phi^{*}_M)=1,\quad N(\Phi^M)=0,\\
&& N(\partial_k)=1,\quad N(\partial_0)=0\\
&& N(dx^{\mu})=0.
\end{eqnarray}
$N$ counts the number of spatial derivatives as well as the antifields
(with equal weight given to each).
According to this degree, $\delta$ decomposes as $\delta_0 + \delta_1$.
The differential $\delta_1$ acts exactly in the same way as the Koszul-Tate
differential
for a system of free 2-forms in 5 dimensions.

We are now able to prove the
\begin{theorem}\label{delta}
$H_i(\delta)=0$ for $i>0$, where $i$ is the antighost number, i.e,
the cohomology of $\delta$ is empty in antighost number strictly
greater than zero.
\end{theorem}
\proof{From \cite{HenneauxKnaepenSchomblond96}, we know that $H_i(\delta_1)=0$.
Let be $a\in\cal{A}$ a $\delta$-closed local function of
antighost number $i>0$. We decompose $a$ according to the degree $N$
\begin{equation}
a=a_1+\ldots+a_m.
\end{equation}
The expansion stops because
$a$ is polynomial in the antifields and the derivatives.
Furthermore, $a_0=0$ because $antigh(a)=i>0$. 
The equation $\delta a=0$ gives in $N$-degree
$m+1$: $\delta_1 a_m=0$.
But $H_i(\delta_1)=0$, hence $a_m=\delta_1 b_{m-1}$. We can define an $a'$ as
being
\begin{equation}
a'=a-\delta b_{m-1}=a_1+\ldots+a_{m-2}+a'_{m-1},
\end{equation}
with $a'_{m-1}=a_{m-1}-\delta_0 b_{m-1}$.
We can proceed in the same way as before with $a'$, whose component of higher
$N$-degree is
of degree less than $m$. We will then find a new $a'$ of highest degree less
than $m-1$,
and so on, each time lowering the $N$-degree.
After a finite number of steps, we arrive at $a^{'}=a^{'}_1=a-\delta b$.
Then, $\delta a=0$ implies $\delta_1 a'_1=0$.
Hence, $a'_1=\delta_1 b_0=\delta b_0$ because $\delta_0\Phi^M=0$.
In conclusion $a=\delta b$, with $b=b_0+\ldots+b_{m-1}$.
}

Of course, this theorem is really a consequence of general known results on the
cohomology of the Koszul-Tate differential.  It simply confirms, in a sense, 
that we have correctly taken into account all gauge symmetries and 
reducibility identities in constructing the antifield spectrum.

The cohomological space $H^{5,inv}_k(\delta\mid\tilde{d})$ is defined as 
$H^{5}_k(\delta\mid\tilde{d})$ in the space of local spatial forms that
belongs to ${\cal{I}}$, i.e., that are invariant.
We want to compute it for $k$ even and $\neq 0$.
To do this, we will proceed as in the proof of theorem
\ref{delta}.
We first prove the requested result for $\delta_1$;
we then use ``cohomological
perturbation" techniques to extend
the result to $\delta$.
\begin{lemma}\label{delta1}
For $k=2,4,\ldots$
\begin{equation}
H_k^{5,inv}(\delta_1\mid\tilde{d})=0.
\end{equation}
\end{lemma}
Again, this result is simply a particular case of 
more general results, which were
previously known, but for completeness, we prove it here.
\proof{Firstly, the theorem 9.1 of \cite{BarnichBrandtHenneaux94D} says that
for a linear gauge theory of reducibility order $p$ in $n$ dimensions
$H^n_k(\delta\mid d)=0$ for $k>p+2$.
A system of abelian spatial 2-forms in 5 dimensions is a linear gauge theory of
reducibility order 1 (see section \ref{action}),
thus, we can state that $H^5_k(\delta_1\mid\tilde{d})=0$ for $k>3$.

Secondly, the theorem 7.4 of \cite{HenneauxKnaepenSchomblond96} gives here :
$H^5_2(\delta_1\mid\tilde{d})=0$.

Finally, the theorem 10.1 of \cite{HenneauxKnaepenSchomblond96} says that for a
system of
space-time p-form gauge fields of the same degree
$H^n_k(\delta\mid d)\cong H^{n,inv}_k(\delta\mid d)$ for $k>0$. For the system
under consideration here, this can be translated
into: $H^5_k(\delta_1\mid \tilde{d})\cong H^{5,inv}_k(\delta_1\mid \tilde{d})$
for $k>0$.
Putting all these results together completes the proof.
}
Let be $a^5(\chi)$ a local spatial $5$-form in ${\cal{I}}$ of strictly 
positive
and even antighost number, satisfying
\begin{equation}
\delta a^5(\chi) + \tilde{d} b^4(\chi)=0.\label{modd}
\end{equation}
We can decompose $a^5$ and $b^4$ according to the degree $N$
\begin{eqnarray}
&& a^5=a^5_1+\ldots+a^5_n,\\
&& b^4=b^4_1+\ldots+b^4_m.
\end{eqnarray}
$a^5_0=0$ and $b^4_0=0$ because $a^5$ and $b^4$ are of antighost number $>0$.
We can always suppose $m\leq n$ because if $m>n$, (\ref{modd}) gives in
$N$-degree $m+1$: $\tilde{d} b^4_m=0$.
Using the invariant Poincar\'e lemma, this yields $b^4_m=\tilde{d}
c^3_{m-1}$. Hence, $b^4_m$ only contributes
to $b^4$ by a $\tilde{d}$-trivial term which can be eliminated.
Proceeding in the same way until $m=n$, we arrive at the equation
\begin{equation}
\delta_1 a^5_n(\chi) + \tilde{d} b^4_n(\chi)=0.
\label{keykey}
\end{equation}
It has already been noticed above
that the algebra ${\cal{I}}$ without dependence on $dx^0$
is the same as for a system of spatial 2-forms.
We can thus use the lemma \ref{delta1} in (\ref{keykey}) to find that
\begin{eqnarray}
a^5_n(\chi)=\delta_1 e^5_{n-1}(\chi) + \tilde{d} f^4_{n-1}(\chi).
\end{eqnarray}
Therefore, $a^{'5}=a^5-\delta e^5_{n-1} - \tilde{d} f^4_{n-1}$ satisfies the
same properties
as $a^5$, except that its component of highest $N$-degree is of
degree $<n$.
We can now apply the same reasoning as before to $a^{'5}$, and so on, until
we arrive at
\begin{equation}
a^{'5}=a^{'5}_1=a^5-\delta (\sum_{i=1}^{n-1}e^5_i) - \tilde{d}
(\sum_{i=1}^{n-1}f^4_i)
\end{equation}
This leads to
\begin{equation}
a^{'5}_1=\delta_1 e^5_0(\chi) + \tilde{d} f^4_0(\chi).
\end{equation}
But $\delta_1 e^5_0=\delta e^5_0$ because $\delta_0\Phi^M=0$.
Eventually, we have $a^5=\delta e^5(\chi) + \tilde{d} f^4(\chi)$,
with $e^5=\sum\limits_{i=0}^{n-1}e^5_i$ and $f^4=\sum\limits_{i=0}^{n-1}f^4_i$.
This gives the awaited theorem:
\begin{theorem}\label{deltadinv}
For $k=2,4,\ldots$
\begin{equation}
H_k^{5,inv}(\delta\mid\tilde{d})=0.
\end{equation}
\end{theorem}

\section{Decomposition of the Wess-Zumino equation}
\setcounter{equation}{0}
\setcounter{theorem}{0}
\setcounter{lemma}{0}

We now have all the necessary tools to
solve the Wess-Zumino consistency condition
that controls the consistent deformations (to first-order) of
the action,
\begin{equation}
sa^6 + db^5 = 0,
\label{WZ}
\end{equation}
where $a^6$ and $b^5$ are local forms of respective form degrees 6 and 5, and
ghost number 0 and 1.
These forms are defined up to the following allowed redefinitions
\begin{eqnarray}
&& a^6\rightarrow a^6+sf^6+dg^5\label{redef1}\\
&& b^5\rightarrow b^5+sg^5+dh^4,
\label{redef2}
\end{eqnarray}
which preserve (\ref{WZ}).
We can decompose $a^6$ and $b^5$ according to antighost number, which gives
\begin{eqnarray}
a^6&=&a^6_0+ \ldots +a^6_k,\\
b^5&=&b^5_0+\ldots+b^5_q,
\end{eqnarray}
with $a^6_k\neq 0$.

We suppose $k>0$ and we will show that $a^6_k$ can be eliminated if we redefine
$a^6$
in an appropriate way.
In antighost number $k$, the equation (\ref{WZ}) just reads
\begin{equation}
\gamma a^6_k + db^5_k=0.
\label{Hgamma}
\end{equation}
We can always assume $k\geq q$ because if $q>k$,
the equation (\ref{WZ}) gives in highest antighost number $db^5_q=0$.
Using the algebraic Poincar\'e lemma, we find that $b^5_q=dc^4_q$.
Hence, we can remove the component $b^5_q$ up to a $d$-trivial redefinition of
$b^5$.

{ } From the theorems \ref{gamma} and \ref{gammad'}, we know that Eq.
(\ref{Hgamma}) implies
\begin{equation}
a^6_k=\sum_I P_I(\chi)\omega^I+\gamma f^6_k + dg^5_k.
\end{equation}
The $\gamma$ modulo $d$ trivial part of $a^6_k$ can be eliminated by 
redefining
$a^6$ in the following way
\begin{equation}
a^6\rightarrow a^6-sf^6_k-dg^5_k.
\end{equation}
We notice that $H^{6,0}_k(\gamma)$
is non trivial only in even antighost number $k$ (because $\eta$ is of
pureghost number 2).
This implies that we can assume $k$ to be even.

The Wess-Zumino consistency condition in antighost number $k-1$ is
\begin{equation}
\gamma a^6_{k-1}+\delta a^6_k+db^5_{k-1}=0.
\label{WZ2}
\end{equation}
The term $b^5_{k-1}$ is invariant because (\ref{WZ2}) implies $d(\gamma
b^5_{k-1})=0$.
Therefore, the algebraic Poincar\'e lemma gives $\gamma b^5_{k-1} +
dc^4_{k-1}=0$ because $k>1$.  From
>From the theorem \ref{gammad'} we know that we can suppose
$\gamma b^5_{k-1}=0$ without affecting $a^6$.
Furthermore, if $b^5_{k-1}=\gamma c^5_{k-1}$ we can eliminate $b^5_{k-1}$ by
redefining $b^5$
in the following way: $b^5\rightarrow b^5-sc^5_{k-1}$, which does not modify
$a^6_k$.

Therefore, we can assume 
\begin{eqnarray}
a^6_k&=&\sum_Idx^0\tilde{P}^5_I\omega^I,\label{dring1}\\
b^5_{k-1}&=&\sum_I(\tilde{Q}^5_I+dx^0\tilde{R}^4_I)\omega^I.
\label{dring2}
\end{eqnarray}
The $\tilde{P}^5_I$, $\tilde{Q}^5_I$, and $\tilde{R}^4_I$
are local spatial forms belonging to ${\cal{I}}$.

Inserting (\ref{dring1}) and (\ref{dring2}) in (\ref{WZ2}), we find
\begin{eqnarray}
\gamma a^6_{k-1}
&=&\sum_I\{-\tilde{d}\tilde{Q}^5_I\omega^I
-\gamma[(\tilde{Q}^5_I+dx^0\tilde{R}^4_I)\hat{\omega}^I]\\
&&+dx^0[(\delta\tilde{P}^5_I+\tilde{d}\tilde{R}^4_I
-\partial_0\tilde{Q}^5_I)\omega^I
-\tilde{Q}^5_I\partial_0\omega^I]\},
\end{eqnarray}
with $\tilde{d}\omega^I=\gamma\hat{\omega}^I$. This implies that
\begin{equation}
\sum_I[(\delta\tilde{P}^5_I+\tilde{d}\tilde{R}^4_I
-\partial_0\tilde{Q}^5_I)\omega^I
-\tilde{Q}^5_I\partial_0\omega^I]=\gamma\beta.
\end{equation}
If we analyse this equation in the same way as the equation (\ref{back2}),
we can prove that $\tilde{Q}^5_I=\delta\tilde{P}^5_I+\tilde{d}\tilde{R}^4_I$
(or simply vanishes).
Inserting these equations in (\ref{dring2}), we find that $b^5_{k-1}$ is of the
form
\begin{equation}
b^5_{k-1}=\delta c^5_{k}+de^4_{k-1}+\gamma f^5_{k-1}
+dx^0\sum_I\tilde{R}^{'4}_I(\chi)\omega^I,
\end{equation}
where $c^5_{k}$ and $e^4_{k-1}$ belong to $H(\gamma)$.
In conclusion, we can eliminate $\tilde{Q}^5_I$ from $b^5_{k-1}$ 
by redefining
$a^6$
and $b^5$ in the following way
\begin{eqnarray}
&& a^6\rightarrow a^6-d(c^5_k+f^5_{k-1}),\\
&& b^5\rightarrow b^5-s(c^5_k+f^5_{k-1})-de^4_{k-1},
\end{eqnarray}
which does not affect the 
condition $\gamma a^6_k=0$, because $\gamma c^5_{k}=0$.

Therefore, we can finally assume
\begin{equation}
a^6_k=\sum_Idx^0\tilde{P}^5_I(\chi)\omega^I,\quad
b^5_{k-1}=\sum_I dx^0\tilde{R}^4_I(\chi)\omega^I.
\end{equation}
The equation (\ref{WZ2}) becomes
\begin{equation}
\gamma a^{'}_{k-1}+dx^0\sum_I(\delta
\tilde{P}^5_I(\chi)+\tilde{d}\tilde{R}^4_I(\chi))\omega^I=0,
\end{equation}
which implies that $\delta \tilde{P}^5_I(\chi)+\tilde{d}\tilde{R}^4_I(\chi)=0$.
We know that we are in even antighost number, thus we can use the theorem
\ref{deltadinv}
to find that $\tilde{P}^5_I=\delta
\tilde{S}^5_I(\chi)+\tilde{d}\tilde{T}^4_I(\chi)$. Hence,
\begin{equation}
a^6_k = sf^6_{k+1}+dg^5_k+\gamma h^6_k,
\end{equation}
where we have defined
\begin{eqnarray}
&& f^6_{k+1}=-dx^0\sum_I\tilde{S}^5_I\omega^I,\quad g^5_k=-dx^0\sum_I
\tilde{T}^4_I\omega^I,\\
&& h^6_k=dx^0\sum_I \tilde{T}^4_I\hat{\omega}^I,\quad
\tilde{d}\omega^I=\gamma\hat{\omega}^I.
\end{eqnarray}
Thus $a^6_k$ can be completely eliminated by redefing $a^6$ as
\begin{equation}
a^{'6}=a^6-s(f^6_{k+1}+ h^6_k)-dg^5_k,
\end{equation}
which only affects the components of antighost number
$<k$.  Repeating the argument at lower antighost numbers
enables one to remove successively $a_{k-1}$, $a_{k-2}$, ...,
up to $a_1$.
This completes the proof of the fact that there is no non
trivial dependence on the antifields for the elements of
$H^{6,0}(s \mid d)$.

For antifield-independent local forms, the
cocycle condition $H^{6,0}(s \mid d)$ reduces to the cocycle condition for
$H^{6,0}(\gamma \mid d)$.  Furthermore, $\gamma$-exact (mod-$d$)
solutions are also $s$-exact.  Thus, we are led to consider
$H^{6,0}(\gamma \mid d)$.  This cohomology  is
given by the theorem \ref{inter}. [The terms in that cohomology
that vanish on-shell are trivial in the $s$-cohomology.]
Thus, the only consistent deformations
of the free action for a system
of abelian chiral $2$-forms are either functions of the curvatures
or of the Chern-Simons type.
In both cases, the integrated deformations are off-shell gauge invariant
and yield no modification of the gauge transformations.

\section{Final comments and conclusions}

We have shown that the most general first-order consistent
deformation of a set of free chiral $2$-forms cannot modify
(non trivially)
the original gauge transformations and a fortiori, their
algebra, which remains abelian.  Thus, there is no room
for a non-abelian, local, generalization of the
theory analogous to
the Yang-Mills construction.

This result holds in fact to all orders, since the allowed
deformations involve the gauge-invariant curvatures or
Chern-Simons terms.  The addition of such terms to the
original action  yields a new action 
which is evidently gauge-invariant under the original gauge transformations
to all orders.

One can show along identical lines
that the rigidity of the gauge symmetries is actually
valid for a set of chiral $2p$-forms in $2p + 2$ dimensions,
for any $p>0$.  If one includes other fields, one may deform
the gauge transformations, but the possibilities are severely
limited \cite{bhs2}. For instance, in 10 dimensions, 
the only couplings of a chiral 4-form to 2-forms are those present in type IIB supergravity.

\vspace{5mm}

\noindent {\bf Acknowledgments}:
X.B. and M.H. are supported in part by the ``Actions de
Recherche Concert{\'e}es" of the ``Direction de la Recherche
Scientifique - Communaut{\'e} Fran{\c c}aise de Belgique", by
IISN - Belgium (convention 4.4505.86) and by
Proyectos FONDECYT 1970151 and 7960001 (Chile).
A.S. is supported in part by the FWO and by the European
Commission TMR programme ERBFMRX-CT96-0045 in which he is
associated to K.\ U.\ Leuven.

\bibliographystyle{BKstyle}

\end{document}